\documentclass[journal abbreviation, manuscript]{copernicus}

\usepackage{tabularx}
\usepackage{cancel}
\usepackage{multirow}
\usepackage{booktabs}
\usepackage{supertabular}
\usepackage{algorithmic}
\usepackage{algorithm}
\usepackage{amsthm}
\usepackage{float}
\usepackage{rotating}

\usepackage{graphicx}
\usepackage{array}

\usepackage{siunitx}

\begin{document}
\nolinenumbers

\title{Investigating impacts of dust events on atmospheric surface temperature in Southwest Asia using AERONET data, satellite recordings, and atmospheric models}


\Author[1][mahsajh@iasbs.ac.ir]{Mahsa}{Jahangiri} 
\Author[1]{Afrooz}{Jouzdani}
\Author[1,2]{Hamid Reza}{Khalesifard}

\affil[1]{Department of Physics, Institute for Advanced Studies in Basic Sciences, No. 444 Prof. Sobouti Blvd.,
Zanjan 4513766731, Iran.}
\affil[2]{Center for Research in Climate Change and Global Warming, Institute for Advanced Studies in Basic Sciences, No. 444 Prof. Sobouti Blvd.,
Zanjan 4513766731, Iran.}




\runningtitle{TEXT}

\runningauthor{TEXT}

\received{}
\pubdiscuss{} 
\revised{}
\accepted{}
\published{}


\firstpage{1}

\maketitle
\begin{abstract}
Dust layers have already been reported to have negative impacts on the radiation budget of the atmosphere. But the questions are: How does the atmospheric surface temperature change during a dust outbreak, and what is its temporal correlation with variations of the dust outbreak strength? We investigated these at selected AERONET sites, including Bahrain, IASBS, Karachi, KAUST Campus, Kuwait University, Lahore, Mezaira, Solar Village, in Southwest Asia, and Dushanbe in Central Asia, using available data from 1998 to 2024. The aerosol optical depth at 870 nm and the temperature recorded at each site are taken as measures of dust outbreak strength and atmospheric surface temperature, respectively. The Hybrid Single-Particle Lagrangian Integrated Trajectory (HYSPLIT) model and the aerosol optical depths recorded by the Moderate Resolution Imaging Spectroradiometers (MODIS) on board the Aqua and Terra satellites are used to specify the sources of the dust outbreaks. Our investigations show that in most cases, the temperature decreases during a dust outbreak, but in a considerable number of cases, the temperature rises. Temperature changes are mostly less than 5 $^\circ$C. We found that a dust outbreak may affect the temperature even up to two days after its highest intensity time. This effect is more profound at sites far from large dust sources, such as IASBS in northwest Iran. For sites that are located on either a dust source or very close to it, the temperature and dust optical depth vary almost synchronously.
\end{abstract}

\introduction  
The atmosphere in Southwest Asia is highly affected by dust storms originating from different sources in the region or neighboring areas \citep{prospero2002environmental}. Such storms have considerable impacts on the atmospheric radiation budget. In other words, dust storms affect surface temperature by blocking solar radiation \citep{stocker2014climate, jouzdani2024investigating}.
Modulation of solar radiation by atmospheric aerosols, including dust, is a critical factor in the climate system \citep{wang2009particulate, nabat2020modulation}. Dust particles, when advected by wind, can cover vast regional to intercontinental distances under suitable meteorological conditions, consequently affecting air quality far from their point of origin \citep{he2024air}. The dispersion of aerosols is influenced by weather conditions, topography, and vegetation \citep{engelstaedter2006north}. These particles can directly alter the atmospheric radiation budget through the absorption and backscattering of sunlight, and indirectly by influencing cloud properties and atmospheric composition \citep{haywood2000estimates}.

About 65 percent of arid lands in Southwest and Central Asia include low topographical areas such as the Karakum Desert in Turkmenistan, Margo Desert and Registan in Afghanistan, Dasht-e Lut and Dasht-e Kavir deserts in Iran, and sand valleys in the Makran mountains along the northern coast of the Arabian Sea. These regions either host active dust sources or have a high potential to become such sources in the future \citep{rashki2013dryness}.
Southwest Asia's alluvial plains and deserts contribute to nearly 20 percent of the global total dust emission. The maximum dust emission rate in this area is approximately equal to that of the Bodele Depression and central Sahara dust sources in Africa \citep{ginoux2012global}.
The situation has been exacerbated by prolonged droughts in Southwest Asia, particularly during $1998-2001$ and $2007-2009$. These droughts have increased the frequency and intensity of dust outbreaks in the region, affecting various countries including Syria, Iraq, Kuwait, Turkey, Afghanistan, Pakistan, Iran, and the Arabian Peninsula \citep{bayat2021characterization}.

Dust events in Southwest Asia have significant regional and seasonal variations. The periods of these events vary across different regions throughout the year, primarily influenced by synoptic conditions \citep{rezazadeh2013climatology}. In addition, the composition and size of dust particles differ depending on their source and the distance of the observation site from the source \citep{masoumi2013retrieval}.

In this work, we investigate the impact of dust events on atmospheric surface temperature in Southwest and Central Asia, using their corresponding Aerosol Optical Depth (AOD) and Angstrom Exponent (AE). These parameters were obtained from nine sites of the Aerosol Robotic Network (AERONET) over the region. We also examined the correlation between the AODs and the temperature. The temporal correlation of dust AOD and surface temperature was investigated in different delay times, i.e., up to five days before and after the dust event in daily steps. Also, utilizing the AERONET recordings, we investigated the variations of dust activities along the year for the whole available data over each site. We tried to specify the origins of detected dust events over the sites using the Hybrid Single-Particle Lagrangian Integrated Trajectory (HYSPLIT) model and the recorded AOD data by the Moderate Resolution Imaging Spectroradiometer (MODIS) onboard Aqua and Terra satellites. 

The rest of this paper is organized as follows. Section \ref{section: Climatology} presents a climatological overview of the study region, including geography, annual precipitation conditions, and temperature variations. Section \ref{section: Instrumentation and data} describes the instrumentation and datasets used in this study. Section \ref{section: Results and discussions} provides the results of our analysis and discussions of the findings. Finally, the conclusions of this work are summarized in Section \ref{section: conclusions}.

\section{Climatology}
\label{section: Climatology}
\begin{figure}[h]
    \centering
    \includegraphics[width=0.8\linewidth]{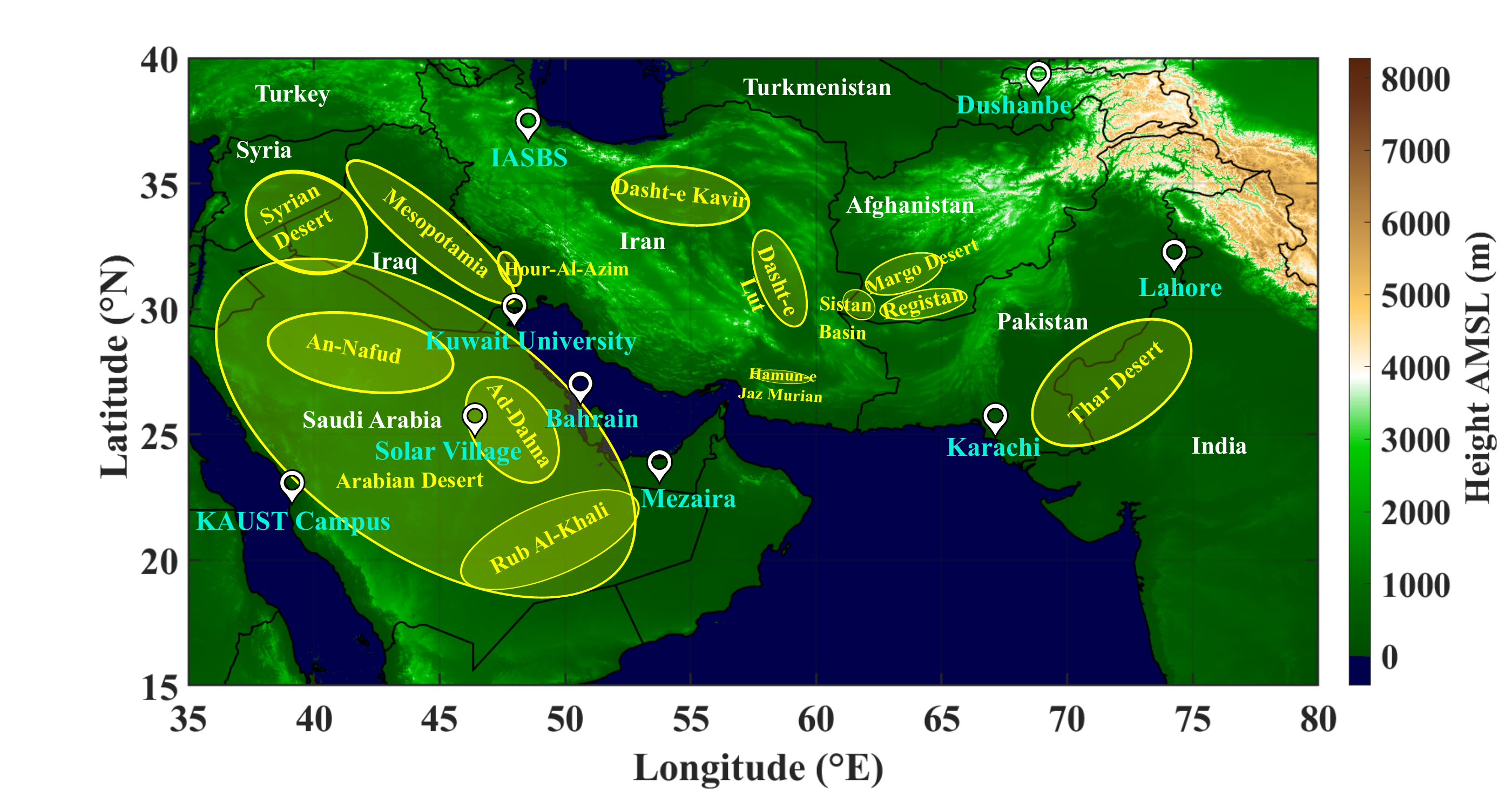}
    \caption{Topography map of the investigation region, locations of the nine selected AERONET sites (white balloons), and main dust sources (yellow ovals)}
    \label{fig:fig1}
\end{figure}

The topography of the region, the geographic locations of the nine selected AERONET sites (white balloons), and the main dust sources (yellow ovals) in Southwest Asia are shown in Fig.~\ref{fig:fig1}. The climates of the selected stations are quite different. Dushanbe at an average elevation of 821 m above mean sea level (AMSL) in Tajikistan, is close to the Pamir Mountains. IASBS (1805 m AMSL) in Northwest Iran, located in the region where the Zagros and Alborz mountains join each other. Lahore, with a moderate elevation (209 m AMSL) in Pakistan, is in the north of the Indo-Gangetic plain. Mezaira (201 m AMSL) in the United Arab Emirates is located north of Rub al-Khali, a desert region. Kuwait University (42 m AMSL), KAUST Campus (11.2 m AMSL) in Saudi Arabia, and Karachi (49 m AMSL) are located in coastal areas. Solar Village, at a considerably high altitude (764 m AMSL), is located in a very dry region. Finally, Bahrain is located on an island at an elevation close to sea level (25 m AMSL). Table~\ref{tab:table1} provides a concise overview of the nine AERONET sites, illustrates their geographical coordinates, elevation, periods of available data, and the number of detected dust events.

\begin{figure}[h]
    \centering
    \includegraphics[width=0.8\linewidth]{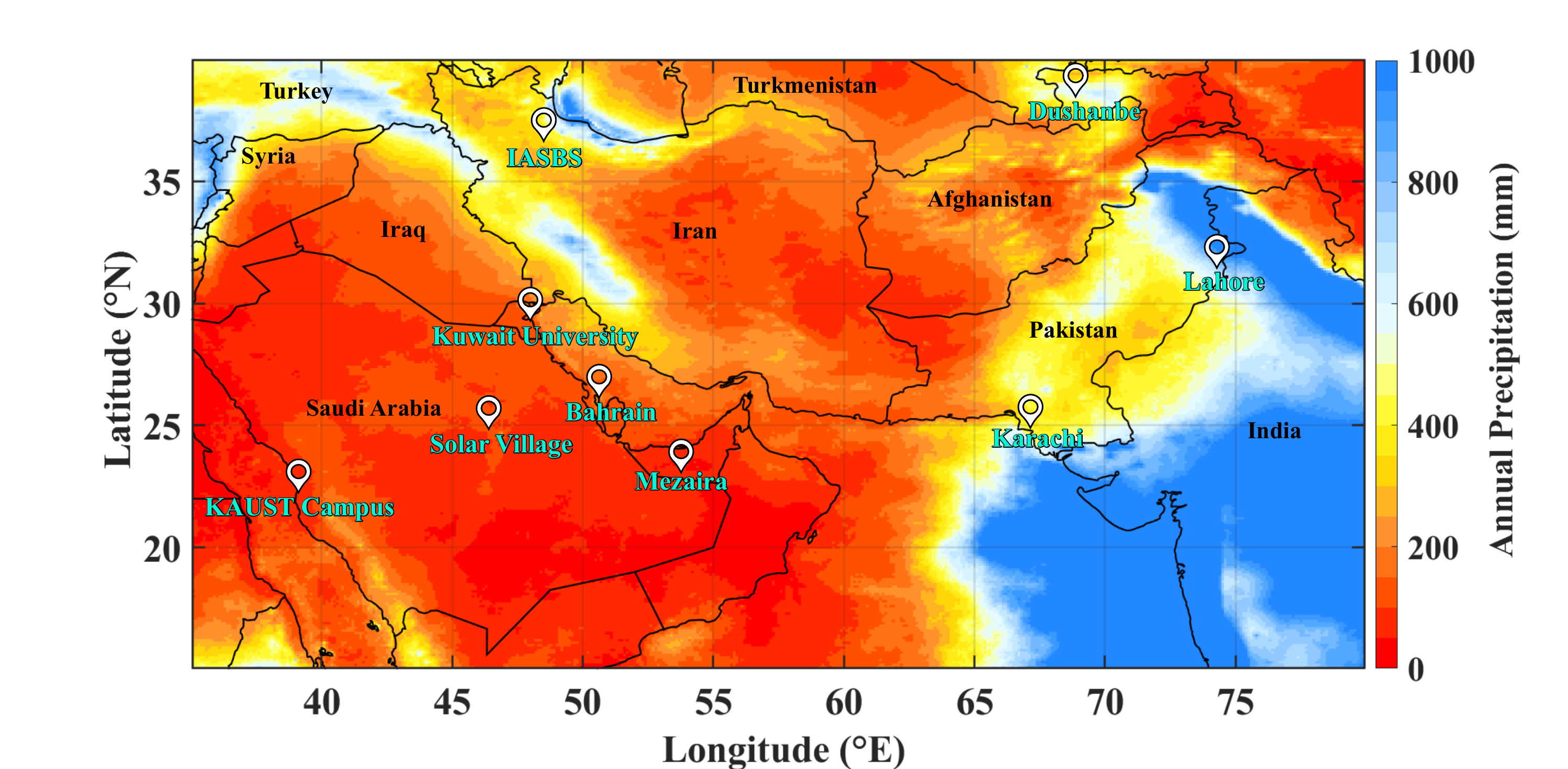}
    \caption{Five-year average precipitation ($2019-2023$) in the region obtained from the NASA GPM IMERG final run daily product~\citep{GPM_IMERG_V07}}
    \label{fig:fig2}
\end{figure}

Figure~\ref{fig:fig2} is a five-year average precipitation ($2019-2023$) in the region obtained from the NASA GPM IMERG final run daily product~\citep{GPM_IMERG_V07}, accessed and analyzed using the Giovanni online data system developed by NASA GES DISC. As Fig.~\ref{fig:fig2} shows, annual precipitation is quite different in the AERONET sites. Lahore and Dushanbe experience the highest precipitation rate, exceeding $600~\mathrm{mm~yr^{-1}}$. Bahrain, Karachi, and IASBS receive moderate rain ($200-400~\mathrm{mm~yr^{-1}}$). Solar Village and Kuwait University are in regions with around $100-150$ mm of annual rainfall, but precipitation at the KAUST Campus and Mezaira sites is often less than $100~\mathrm{mm~yr^{-1}}$.

\begin{figure}[h]
    \centering
    \includegraphics[width=0.8\linewidth]{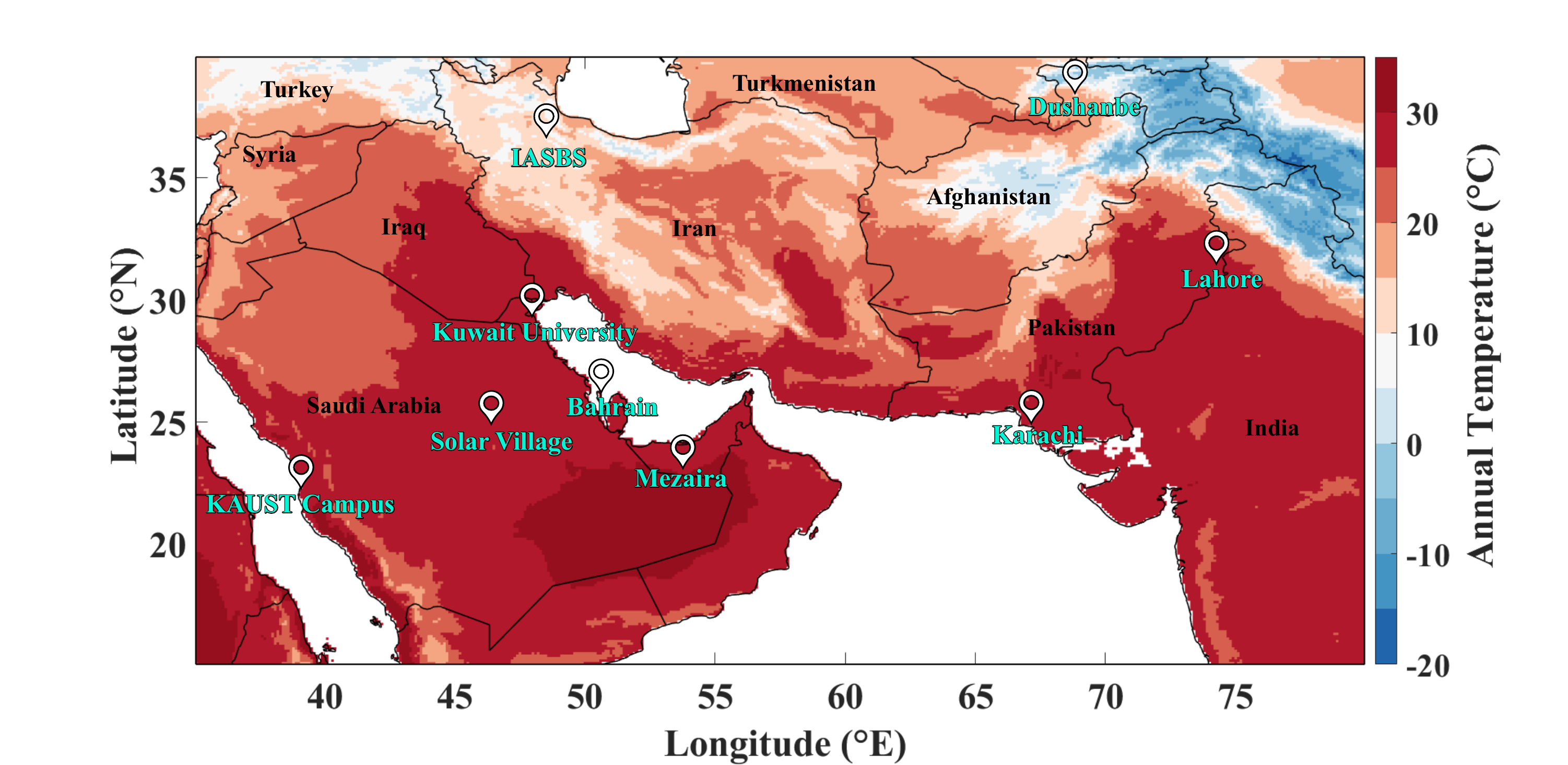}
    \caption{Five-year average surface air temperature ($2019-2023$) in the region obtained from the NASA FLDAS Noah L4 monthly~\citep{FLDAS_NOAH_CHIRPS_v001}}
    \label{fig:fig3}
\end{figure}

The five-year average temperature is obtained from the NASA FLDAS Noah L4 monthly~\citep{FLDAS_NOAH_CHIRPS_v001}, averaged over the years $2019-2023$ (Fig.~\ref{fig:fig3}). Mezaira has the highest average temperature of approximately 30 $^\circ$C. Karachi, Lahore, Solar Village, KAUST Campus, Kuwait University, and Bahrain are cooler and experience annual surface temperatures of $25-30$ $^\circ$C. Dushanbe and IASBS have significantly lower temperatures ($10-15$ $^\circ$C).

\section{Instrumentation and data}
\label{section: Instrumentation and data}
In this section, the instrumentation and data used in this work are presented. These include the AERONET and its recordings at the selected sites (Fig.~\ref{fig:fig1}), MODIS AODs at 550 nm, and the HYSPLIT model.

\subsection{AERONET}
AERONET is a worldwide distributed network of ground-based sun photometers that delivers detailed, long-term, and globally uniform data on aerosol optical characteristics \citep{holben1998aeronet}. AERONET provides a comprehensive collection of data products, freely accessible through NASA’s Goddard Space Flight Center (GSFC) website. Version 3 AOD data are available in multiple processing stages, including Level 1.0 (unscreened), Level 1.5 (cloud-screened and quality-controlled), and Level 2.0 (fully quality-assured) \citep{giles2019advancements}. AERONET employs advanced inversion algorithms to derive various aerosol properties, including AOD, AE, surface air temperature, and many other optical parameters of the atmosphere \citep{dubovik2000accuracy}. The optical depth of a medium determines the amount of light intensity reduction as it passes through the medium. The AE is an intensive physical parameter of atmospheric aerosols that depends on particle size \citep{bn1986characteristics}. From the recordings at the AERONET sites, we used the AOD at 870 nm, the AE (calculated from measurements at 440 nm and 870 nm), and the recorded temperatures. These parameters were obtained from the cloud-screened AERONET Level 1.5 dataset. The selected AERONET sites include Bahrain in Bahrain, Dushanbe in Tajikistan, the Institute for Advanced Studies in Basic Sciences (IASBS) in Iran, Karachi and Lahore in Pakistan, the King Abdullah University of Science and Technology (KAUST) Campus and Solar Village in Saudi Arabia, Kuwait University in Kuwait, and Mezaira in the United Arab Emirates. Table~\ref{tab:table1} provides the geographical locations, the duration of the available data, and the number of dust events detected at each AERONET site. The dust events are specified based on their corresponding daily averaged AERONET-reported AODs and AEs. From now on, we call these two parameters $\tau_i$ and $\alpha_i$. In this work, whenever $\tau >0.35$ and $\alpha <0.5$, we considered the case as a dust event.

\begin{table}[]
\centering
\caption{Geographical locations, elevations, available data, and number of detected dust events (AOD $>$ 0.35 and AE $<$ 0.5 ) for
nine AERONET sites. Acceptable dusty days are those with cloud coverage less than 0.25.}
\label{tab:table1}
\begin{tabular}{l l l lll llll}
\toprule
\multicolumn{1}{c}{\multirow{3}{*}{Site/ Country}} & \multicolumn{1}{c}{\multirow{3}{*}{Location}} & \multicolumn{1}{c}{\multirow{3}{*}{\begin{tabular}[c]{@{}c@{}}Elevation \\ AMSL {[}m{]}\end{tabular}}} & \multicolumn{3}{c}{Available Data} & \multicolumn{4}{l}{Detected Dust Events} \\ \cmidrule{4-10} 
\multicolumn{1}{c}{} & \multicolumn{1}{c}{} & \multicolumn{1}{c}{} & \multicolumn{1}{c}{\multirow{2}{*}{From}} & \multicolumn{1}{c}{\multirow{2}{*}{To}} & \multicolumn{1}{c}{\multirow{2}{*}{\begin{tabular}[c]{@{}c@{}}Total\\ (Days)\end{tabular}}} & \multicolumn{2}{c}{Total} & \multicolumn{2}{c}{Acceptable} \\ \cmidrule{7-10} \multicolumn{1}{c}{} & \multicolumn{1}{c}{} & \multicolumn{1}{c}{} & \multicolumn{1}{c}{} & \multicolumn{1}{c}{} & \multicolumn{1}{c}{} & \multicolumn{1}{c}{Days} & \multicolumn{1}{c}{\%} & \multicolumn{1}{c}{Days} & \multicolumn{1}{c}{\%} \\ \midrule

Bahrain/ Bahrain               & ($26.2^{\circ} N, 50.6^{\circ} E$)  & 25    & \multicolumn{1}{l}{1998}   & \multicolumn{1}{l}{2011}   & 1125  & \multicolumn{1}{l}{141}  & \multicolumn{1}{l}{13}  & \multicolumn{1}{l}{20}   & 2 \\
Dushanbe/ Tajikistan           & ($38.6^{\circ} N, 68.9^{\circ} E$)  & 821   & \multicolumn{1}{l}{2010}   & \multicolumn{1}{l}{2024}   & 2817  & \multicolumn{1}{l}{212}  & \multicolumn{1}{l}{8}  & \multicolumn{1}{l}{35}   & 1 \\
IASBS/ Iran                    & ($36.7^{\circ} N, 48.5^{\circ} E$)  & 1805  & \multicolumn{1}{l}{2007}   & \multicolumn{1}{l}{2021}   & 1560  & \multicolumn{1}{l}{70}   & \multicolumn{1}{l}{5}  & \multicolumn{1}{l}{13}   & 1 \\
Karachi/ Pakistan              & ($24.9^{\circ} N, 67.1^{\circ} E$)  & 49    & \multicolumn{1}{l}{2010}   & \multicolumn{1}{l}{2020}   & 1852  & \multicolumn{1}{l}{439}  & \multicolumn{1}{l}{23} & \multicolumn{1}{l}{58}   & 3 \\
KAUST Campus/ Saudi Arabia     & ($22.3^{\circ} N, 39.1^{\circ} E$)  & 11.2  & \multicolumn{1}{l}{2012}   & \multicolumn{1}{l}{2024}   & 3398  & \multicolumn{1}{l}{463}  & \multicolumn{1}{l}{14} & \multicolumn{1}{l}{86}   & 3 \\
Kuwait University/ Kuwait      & ($29.3^{\circ} N, 48.0^{\circ} E$)  & 42    & \multicolumn{1}{l}{2008}   & \multicolumn{1}{l}{2021}   & 3249  & \multicolumn{1}{l}{518}  & \multicolumn{1}{l}{16} & \multicolumn{1}{l}{90}   & 3 \\
Lahore/ Pakistan               & ($31.5^{\circ} N, 74.3^{\circ} E$)  & 209   & \multicolumn{1}{l}{2006}   & \multicolumn{1}{l}{2024}   & 3860  & \multicolumn{1}{l}{317}  & \multicolumn{1}{l}{8}  & \multicolumn{1}{l}{54}   & 1 \\
Mezaira/ United Arab Emirates  & ($23.1^{\circ} N, 53.8^{\circ} E$)  & 201   & \multicolumn{1}{l}{2004}   & \multicolumn{1}{l}{2018}   & 2458  & \multicolumn{1}{l}{477}  & \multicolumn{1}{l}{19} & \multicolumn{1}{l}{71}   & 3 \\
Solar Village/ Saudi Arabia    & ($24.9^{\circ} N, 46.4^{\circ} E$)  & 764   & \multicolumn{1}{l}{1999}   & \multicolumn{1}{l}{2015}   & 4341  & \multicolumn{1}{l}{861}  & \multicolumn{1}{l}{20}  & \multicolumn{1}{l}{94}   & 2 \\ \bottomrule
\end{tabular}
\end{table}

\subsection{MODIS}
MODIS, an instrument onboard NASA’s Aqua and Terra satellites, is designed to capture detailed data on Earth's surface and atmosphere across various spectral bands. These satellites provide near-global daily coverage, enabling continuous monitoring of Earth's environmental and atmospheric conditions \citep{justice1998moderate}. MODIS AODs at 550 nm are used to determine the intensity of dust plumes, and in association with the AERONET data and HYSPLIT model outputs, provide information about the dust events' source locations. The AODs (550 nm) are obtained from the MODIS data level 3, collection 6.1, Dark Target product, available from both Terra and Aqua satellites, with $1^{\circ} \times 1^{\circ}$ of surface resolution \citep{MODIS_L3_08_M3}. 

\subsection{HYSPLIT}
HYSPLIT model, developed by NOAA’s Air Resources Laboratory, is a powerful tool to model the transport, dispersion, and deposition of atmospheric particles and gases \citep{stein2015noaa}. The HYSPLIT is used for backward trajectory analysis of the observed dust events at the selected AERONET sites.

\section{Results and discussions}
\label{section: Results and discussions}
\begin{figure}[ht]
    \centering
    \includegraphics[width=0.35\linewidth]{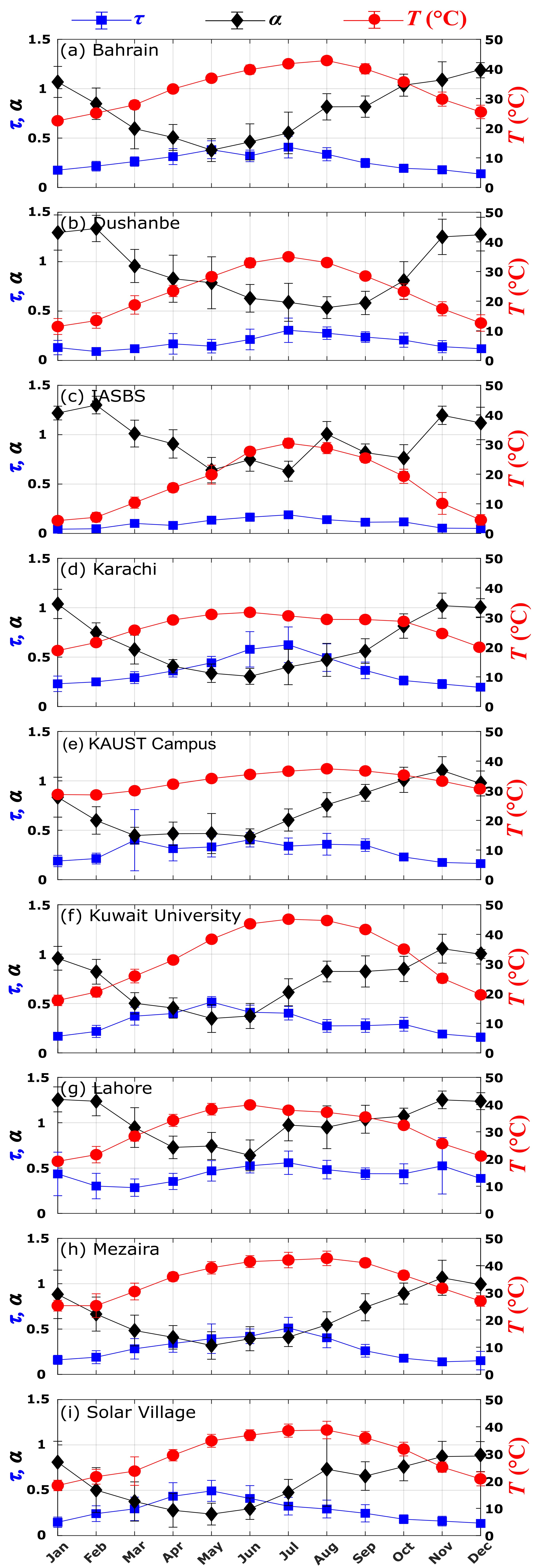}
    \caption{(a)-(i) Monthly variation of $\tau$, $\alpha$, and temperature at nine AERONET sites: Bahrain, Dushanbe, IASBS, Karachi, KAUST Campus, Kuwait University, Lahore, Mezaira, and Solar Village.}
    \label{fig:fig04}
\end{figure}

Monthly average values of $\tau$, $\alpha$, and Midday Temperature (MDT) (average temperature from 09:30 to 15:30 for each day and then throughout the month) for the nine selected AERONET sites are shown in Figs.~\ref{fig:fig04}a–i (connected points blue squares, black diamonds, and red circles, respectively). The connection lines between the points in Figs.~\ref{fig:fig04} are just for clarity, and the error bars are standard deviations of the recorded data for each point. Referring to Fig.~\ref{fig:fig04}, it can be found that almost all sites experienced the most dusty days from May to September. Between the nine sites, IASBS is the least affected site by dust, Kuwait University experienced the highest MDT in July ($\sim 45~^{\circ}\mathrm{C}$) (Fig.~\ref{fig:fig04}f), and it seems that most intense dust events ($\tau$ $\sim 0.63$, $\alpha$ $\sim 0.40$) occurred in Karachi during July (Fig.~\ref{fig:fig04}d).

Knowing that $\tau$ impacts the atmospheric radiation budget, investigating the atmospheric surface temperature and $\tau$ during a dust outbreak, the temporal correlation between these two parameters can be investigated. The temporal correlation ($r_m(\tau,T)$) between $\tau$ and the atmospheric surface temperature $T$ can be written as: 
\begin{equation}
\label{eq:corr}
r_m(\tau,T) = \frac{\sum_{i=-N}^{N-m} (\tau_i - \overline{\tau})(T_{i+m} - \overline{T})}{\sqrt{\left[\sum_{i=-N}^{N} (\tau_i - \overline{\tau})^2\right] \left[\sum_{j=-N}^{N} (T_j - \overline{T})^2\right]}}.
\end{equation}
In Eq.~(\ref{eq:corr}), $i$ is an index to count days and $m$ is the time delay between $\tau$ and the temperature measurements in days. The correlation is calculated for $2N+1$ days, where the most intense dusty day is on day $i=0$. In most cases, we took $N=5$ for the persistent dusty days. $T$ is the MDT obtained from the AERONET recordings for each day. $\overline{T}$ and $\overline{\tau}$ are average values of $T$ and $\tau$ during the $2N+1$ days.
  
\begin{figure}[H]
    \centering
    \includegraphics[width=0.5\linewidth]{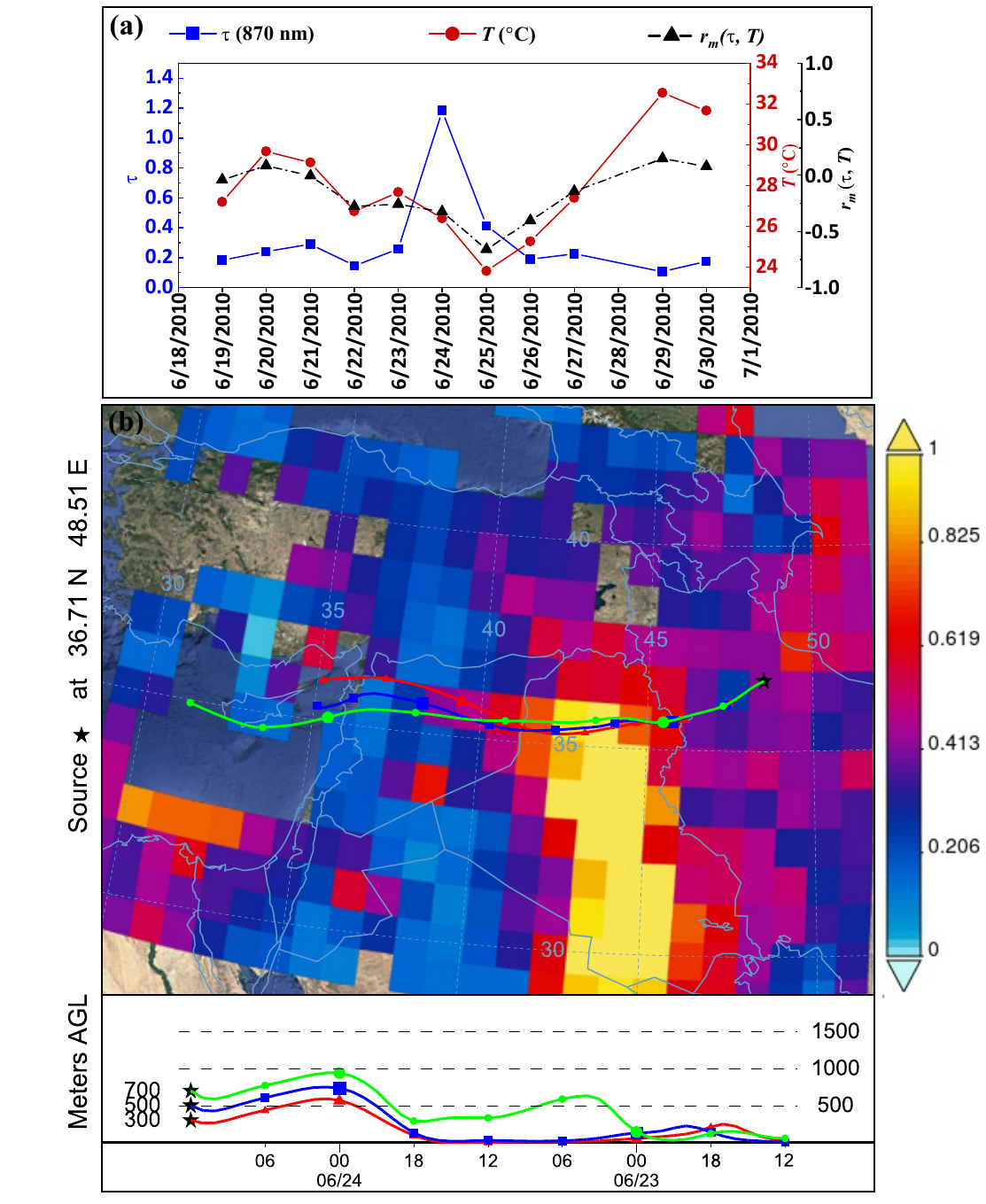}
    \caption{(a) Time series of $\tau$ at 870 nm, temperature (${^\circ} \mathrm{C}$), and correlation coefficient from $19-30$ June 2010; (b) 48 h HYSPLIT backward trajectories ending at 300 m, 500 m, and 700 m AGL at 12:00 UTC on 24 June 2010, overlaid on the MODIS AOD at 550 nm on 23 June 2010, IASBS site.}
    \label{fig:fig5}
\end{figure}

Our investigations reveal that the atmospheric surface temperature changes during and after a dust event. In most cases, changes in $T$ (mainly a decrease) reach to a maximum value a couple of days after the event. In other words, there is a delay between the increase in $\tau$ and the decrease in $T$ after an intense dusty day. As a typical case, Fig.~\ref{fig:fig5}a shows variations of $\tau$ and $T$ at the IASBS site from $19-30$ June 2010. In this case, an event started on 23 June, the highest value of $\tau$ is on 24 June ($\sim 1.2$), and on 25 June (one day after the most intense dusty day), $T$ decreased $\sim 3~^{\circ}\mathrm{C}$. In addition, the minimum temporal correlation between $\tau$ and $T$ is on June 25 (black dashed-dotted line). To quantify the temperature changes, the average value of $T$ for the last two days before the event is subtracted from its minimum value after the event. To specify the dust source of this event, HYSPLIT backward trajectories (starting at the IASBS site on 24 June) are overlaid on the MODIS AODs (at 550 nm) over the region in Fig.~\ref{fig:fig5}b. The figure shows, in less than 24 hours, the trajectories ending at 300 m and 500 m Above Ground Level (AGL), reaching the surface in the north of Mesopotamia, where a dense dust plume exists.
  
Figure~\ref{fig:fig5}a depicts the typical behavior of variations in $\tau$ and $T$ for many cases recorded at selected sites. Of course, the changes in $\tau$, $T$, and the delay between these two parameters vary in different cases. We repeated the source identification procedure shown in Fig.~\ref{fig:fig5}b for all dust events recorded at the selected sites. Since the sunphotometers' recordings lack spatial resolution, source identification using this technique is not accurate. However, when a considerable number of events are available, the procedure provides an acceptable estimate of the dust source's impact on a measurement site. 

\begin{figure}[H]
    \centering
    \includegraphics[width=1\linewidth]{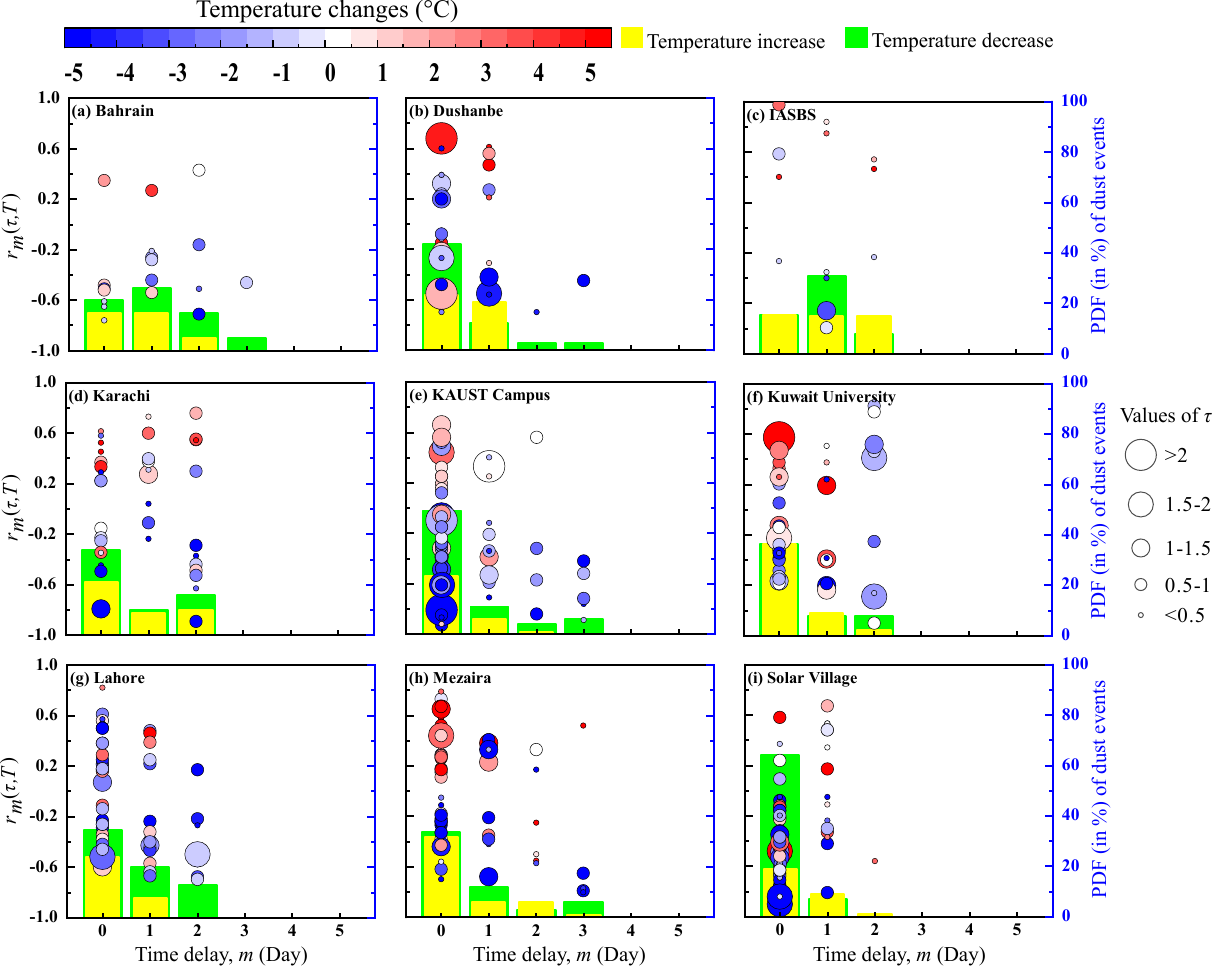}
    \caption{(a)-(i), AODs' ($\tau$) of recorded dust events at nine AERONET sites (colored circles) at different temporal correlation between $\tau$ and $T$, shown by $r_m(\tau,T)$ on the left y-axis and time delays of $m$ days (x-axis). The right y-axis is the Probability Distribution Function (PDF) of the occurred dust events (in $\%$) associated either with a decrease (green bars) or an increase (yellow bars) in $T$.}
    \label{fig:Fig06}
\end{figure}

Figures~\ref{fig:Fig06}a-i depict how different dust events affect the surface temperature at the nine selected sites. In Figs.~\ref{fig:Fig06}, the x axis is the time delay (in days) between the increase in $\tau$ and the changes in $T$and the left y axis is the temporal correlation value $r_m(\tau, T)$ (Eq.~(\ref{eq:corr})) where the recorded $r_m(\tau,T)$ values are shown by colored circles of different sizes. The size and colors of the circles correspond to the values of $\tau$ and the amounts of change in $T$, respectively. The right y-axis in Figs.~\ref{fig:Fig06} corresponds to the probability distribution function (PDF) of the occurred dust events (in $\%$) associated either with a decrease (green bars) or an increase (yellow bars) in $T$.

We have used the HYSPLIT backward trajectory model and the MODIS AODs to specify the sources of the recorded dust events at the sites (almost the same process as shown in Fig.~\ref{fig:fig5}b). Table~\ref{tab:table2} presents the contributions of different dust sources to the dust activities recorded at the nine selected AERONET sites. Referring to column 8 of Table~\ref{tab:table1}, it can be found that Karachi, Solar Village, Mezaira, Kuwait University, KAUST campus, and Bahrain experienced the largest number of dusty days based on their available data (that is, 23$\%$, 20$\%$, 19$\%$, 16$\%$, 14$\%$, and 13$\%$ of available data, respectively). Table~\ref{tab:table2} shows that the Arabian desert, which includes the an-Nafud, ad-Dahna, and Rub al-Khali deserts, is the most influential source in the region and has considerable impacts on Bahrain, Karachi, KAUST Campus, Mezaira, Kuwait University, Solar Village, and even IASBS sites. Mesopotamia impacts the IASBS, Kuwait, and Solar Village sites very frequently. Kyzylkum Desert is the main dust source of recorded events in Dushabe, and the Thar Desert is an active source for Lahore.
\begin{table}[H]
\centering
\caption{Number of dust events originated from the specified dust sources and affected the sites for the time periods mentioned in Table~\ref{tab:table1}}
\label{tab:table2}
\begin{tabular}{lccccccccccc}
\toprule
& \multicolumn{9}{c}{Sites} \\
\cmidrule(lr){2-10}
{Dust sources} & \raisebox{-1.25 cm}{\rotatebox{90}{Bahrain}} & \raisebox{-1.25 cm}{\rotatebox{90}{Dushanbe}} & \raisebox{-1.25 cm}{\rotatebox{90}{IASBS}} & \raisebox{-1.25 cm}{\rotatebox{90}{Karachi}} & \raisebox{-1.25 cm}{\rotatebox{90}{KAUST Campus}} & \raisebox{-1.25 cm}{\rotatebox{90}{Kuwait University}} & \raisebox{-1.25 cm}{\rotatebox{90}{Lahore}} & \raisebox{-1.25 cm}{\rotatebox{90}{Mezaira}} & \raisebox{-1.25 cm}{\rotatebox{90}{Solar Village}} & {Total}\\ \midrule
Sahara Desert      &     &     &     &     & 16  & 6   &     &     &     & 22  \\ 
Arabian Desert     & 11  &     & 6   & 7   & 53  & 43  &     & 49  & 48  & 217 \\ 
Mesopotamia        & 7   &     & 7   &     & 7   & 16  &     & 22  & 30  & 89  \\ 
Hour-Al-Azim       &     &     &     &     &     &     &     &     & 16  & 16  \\ 
Syrian Desert      &     &     &     &     & 6   & 13  &     &     &     & 19  \\ 
Dasht-e Kavir      &     & 1   &     &     &     & 2   &     &     &     & 3   \\ 
Dasht-e Lut        &     &     &     & 4   &     &     & 1   &     &     & 5   \\ 
Hamun-e Jaz Murian &     &     &     & 8   &     &     &     &     &     & 8   \\ 
Sistan Basin       &     &     &     & 6   &     &     &     &     &     & 6   \\ 
Margo Desert       &     &     &     & 1   &     &     & 1   &     &     & 2   \\ 
Registan Desert    &     &     &     & 8   &     &     & 6   &     &     & 14  \\ 
Thar Desert        &     &     &     & 9   &     &     & 45  &     &     & 54  \\ 
Kyzylkum Desert    &     & 21  &     &     &     &     &     &     &     & 21  \\ 
Aralkum Desert     &     & 8   &     &     &     &     &     &     &     & 8   \\ 
Karakum Desert     &     & 5   &     &     &     &     &     &     &     & 5   \\ 
Taklamakan Desert  &     &     &     &     &     &     & 1   &     &     & 1   \\ 
NA                 & 2   &     &     & 1   & 4   &     &     &     &     & 7   \\ \bottomrule
\end{tabular}
\end{table}

\conclusions  
\label{section: conclusions}
We investigated dust activities and their impacts on atmospheric surface temperature at nine AERONET sites in southwest Asia (Fig.~\ref{fig:fig1}). Almost all sites, except Dushanbe and Lahore, are located in arid and semi-arid regions where annual precipitation is about 400 mm or less (Fig.~\ref{fig:fig2}), and the annual surface temperature is higher than $20~^{\circ}\mathrm{C}$ (except IASBS and Lahore, Fig.~\ref{fig:fig3}). However, two more points can be concluded from Figs.~\ref{fig:fig1} and \ref{fig:Fig06}: the IASBS site is quite far from the main dust sources; therefore, dust plumes reaching this site are expected to have considerable shares of fine dust particles. Perhaps this is one of the reasons that the particles stayed longer in the atmosphere, and as a result, the maximum value of $r_m(\tau,T)$ occurs mainly two days after the most dusty day. However, at sites such as the KAUST Campus, Solar Village Kuwait University, and Mezaira that are located either inside a dust source or in its vicinity, there is almost no delay between the appearance of dust layers over the site and changes in the atmospheric surface temperature.

Referring to Fig.~\ref{fig:Fig06}, it can be observed that, in most cases and at all sites, dust events have a negative impact on the atmospheric surface temperature. Meanwhile, the atmospheric surface temperature increased after a considerable number of dust events. It is well known that dust events are initiated by the blowing of high-speed winds over surfaces covered with loose soil \citep{shao2006review}. Such conditions may occur when a low-pressure atmospheric system is active in the region, which in most cases is associated with temperature drops. In addition, dust particles have negative impacts on the atmospheric radiation budget \citep{stocker2014climate}. But this is not the whole story; pressure gradients and even surface heating during daytime can also produce rapidly ascending heated air parcels that eventually form dust plumes \citep{marsham2021dust}. In this work, we didn't go through the details of the dynamics of each of the dust events observed over the selected sites. Therefore, we cannot explain why, in many cases, dust events are associated with an increase in surface temperature. This is the subject of another investigation that we are currently working on it.

\begin{acknowledgements}
The authors are grateful to Ahmad Assar Enayati and Ali Bayat for valuable discussions. This research was facilitated by the availability of AERONET data, provided by NASA's Goddard Space Flight Center (http://aeronet.gsfc.nasa.gov/). We thank the principal investigators, co-investigators, and their staff for establishing and maintaining the eight AERONET sites used in this study: Bahrain, Dushanbe, Karachi, KAUST Campus, Kuwait University, Lahore, Mezaira, and Solar Village. We also acknowledge the Giovanni online data system (https://giovanni.gsfc.nasa.gov/), developed and maintained by NASA's Goddard Earth Sciences Data and Information Services Center (GES DISC), for enabling easy access to Earth science data. Additionally, we thank the NOAA Air Resources Laboratory for providing the HYSPLIT transport and dispersion model (https://www.ready.noaa.gov/), which substantially contributed to the quality of this research. 
\end{acknowledgements}






\bibliographystyle{copernicus}
\bibliography{references}

\begin{thebibliography}{24}
\providecommand{\natexlab}[1]{#1}
\providecommand{\url}[1]{\texttt{#1}}
\providecommand{\urlprefix}{}
\expandafter\ifx\csname urlstyle\endcsname\relax
  \providecommand{\doi}[1]{https://doi.org/\discretionary{}{}{}#1}\else
  \providecommand{\doi}{https://doi.org/\discretionary{}{}{}\begingroup \urlstyle{rm}\Url}\fi

\bibitem[{Bayat and Khalesifard(2021)}]{bayat2021characterization}
Bayat, F. and Khalesifard, H.~R.: Characterization of released dust over open waters in the south of the Iran Plateau based on satellite and ground-based measurements, Atmospheric Pollution Research, 12, 101\,208, 2021.

\bibitem[{BN(1986)}]{bn1986characteristics}
BN, H.: Characteristics of maximum-value composite images from temporal AVHRR data, International Journal of Remote Sensing, 15, 145--161, 1986.

\bibitem[{Dubovik et~al.(2000)Dubovik, Smirnov, Holben, King, Kaufman, Eck, and Slutsker}]{dubovik2000accuracy}
Dubovik, O., Smirnov, A., Holben, B., King, M., Kaufman, Y., Eck, T., and Slutsker, I.: Accuracy assessments of aerosol optical properties retrieved from Aerosol Robotic Network (AERONET) Sun and sky radiance measurements, Journal of Geophysical Research: Atmospheres, 105, 9791--9806, 2000.

\bibitem[{Engelstaedter et~al.(2006)Engelstaedter, Tegen, and Washington}]{engelstaedter2006north}
Engelstaedter, S., Tegen, I., and Washington, R.: North African dust emissions and transport, Earth-Science Reviews, 79, 73--100, 2006.

\bibitem[{Giles et~al.(2019)Giles, Sinyuk, Sorokin, Schafer, Smirnov, Slutsker, Eck, Holben, Lewis, Campbell et~al.}]{giles2019advancements}
Giles, D.~M., Sinyuk, A., Sorokin, M.~G., Schafer, J.~S., Smirnov, A., Slutsker, I., Eck, T.~F., Holben, B.~N., Lewis, J.~R., Campbell, J.~R., et~al.: Advancements in the Aerosol Robotic Network (AERONET) Version 3 database--automated near-real-time quality control algorithm with improved cloud screening for Sun photometer aerosol optical depth (AOD) measurements, Atmospheric Measurement Techniques, 12, 169--209, 2019.

\bibitem[{Ginoux et~al.(2012)Ginoux, Prospero, Gill, Hsu, and Zhao}]{ginoux2012global}
Ginoux, P., Prospero, J.~M., Gill, T.~E., Hsu, N.~C., and Zhao, M.: Global-scale attribution of anthropogenic and natural dust sources and their emission rates based on MODIS Deep Blue aerosol products, Reviews of Geophysics, 50, 2012.

\bibitem[{Haywood and Boucher(2000)}]{haywood2000estimates}
Haywood, J. and Boucher, O.: Estimates of the direct and indirect radiative forcing due to tropospheric aerosols: A review, Reviews of geophysics, 38, 513--543, 2000.

\bibitem[{He et~al.(2024)He, Kumar, Tang, Pfister, Xu, Qian, and Brasseur}]{he2024air}
He, C., Kumar, R., Tang, W., Pfister, G., Xu, Y., Qian, Y., and Brasseur, G.: Air Pollution Interactions with Weather and Climate Extremes: Current Knowledge, Gaps, and Future Directions, Current Pollution Reports, pp. 1--13, 2024.

\bibitem[{Holben et~al.(1998)Holben, Eck, Slutsker, Tanr{\'e}, Buis, Setzer, Vermote, Reagan, Kaufman, Nakajima et~al.}]{holben1998aeronet}
Holben, B.~N., Eck, T.~F., Slutsker, I.~a., Tanr{\'e}, D., Buis, J., Setzer, A., Vermote, E., Reagan, J.~A., Kaufman, Y., Nakajima, T., et~al.: AERONET—A federated instrument network and data archive for aerosol characterization, Remote sensing of environment, 66, 1--16, 1998.

\bibitem[{Huffman et~al.(2023)Huffman, Stocker, Bolvin, Nelkin, and Tan}]{GPM_IMERG_V07}
Huffman, G.~J., Stocker, E.~F., Bolvin, D.~T., Nelkin, E.~J., and Tan, J.: {GPM IMERG Final Precipitation L3 1 day 0.1 degree x 0.1 degree V07}, \doi{10.5067/GPM/IMERGDF/DAY/07}, accessed: 2025-08-03, 2023.

\bibitem[{Jacob and Slinski(2021)}]{FLDAS_NOAH_CHIRPS_v001}
Jacob, J. and Slinski, K.: {FLDAS Noah Land Surface Model L4 Global Monthly 0.1 x 0.1 degree (GDAS and CHIRPS-PRELIM)}, \doi{10.5067/L8GPRQWAWHE3}, accessed: 2025-08-04, 2021.

\bibitem[{Jouzdani et~al.(2024)Jouzdani, Bayat, and Khalesifard}]{jouzdani2024investigating}
Jouzdani, A., Bayat, A., and Khalesifard, H.~R.: Investigating impacts of dust events on surface temperature and atmospheric radiative forcing in Zanjan, Kuwait, and, Karachi, in: E3S Web of Conferences, vol. 575, p. 05004, EDP Sciences, 2024.

\bibitem[{Justice et~al.(1998)Justice, Vermote, Townshend, Defries, Roy, Hall, Salomonson, Privette, Riggs, Strahler et~al.}]{justice1998moderate}
Justice, C.~O., Vermote, E., Townshend, J.~R., Defries, R., Roy, D.~P., Hall, D.~K., Salomonson, V.~V., Privette, J.~L., Riggs, G., Strahler, A., et~al.: The Moderate Resolution Imaging Spectroradiometer (MODIS): Land remote sensing for global change research, IEEE transactions on geoscience and remote sensing, 36, 1228--1249, 1998.

\bibitem[{Marsham and Ryder(2021)}]{marsham2021dust}
Marsham, J.~H. and Ryder, C.~L.: Dust storms and haboobs, Weather, 76, 378--379, 2021.

\bibitem[{Masoumi et~al.(2013)Masoumi, Khalesifard, Bayat, and Moradhaseli}]{masoumi2013retrieval}
Masoumi, A., Khalesifard, H., Bayat, A., and Moradhaseli, R.: Retrieval of aerosol optical and physical properties from ground-based measurements for Zanjan, a city in Northwest Iran, Atmospheric research, 120, 343--355, 2013.

\bibitem[{Nabat et~al.(2020)Nabat, Somot, Cassou, Mallet, Michou, Bouniol, Decharme, Drug{\'e}, Roehrig, and Saint-Martin}]{nabat2020modulation}
Nabat, P., Somot, S., Cassou, C., Mallet, M., Michou, M., Bouniol, D., Decharme, B., Drug{\'e}, T., Roehrig, R., and Saint-Martin, D.: Modulation of radiative aerosols effects by atmospheric circulation over the Euro-Mediterranean region, Atmospheric Chemistry and Physics, 20, 8315--8349, 2020.

\bibitem[{Platnick et~al.(2015)Platnick, Hubanks, Meyer, and King}]{MODIS_L3_08_M3}
Platnick, S., Hubanks, P., Meyer, K., and King, M.~D.: {MODIS Atmosphere L3 Monthly Product (08\_L3) -- Terra and Aqua}, \urlprefix\url{http://dx.doi.org/10.5067/MODIS/MOD08_M3.006 and http://dx.doi.org/10.5067/MODIS/MYD08_M3.006}, accessed: 2025-08-11, 2015.

\bibitem[{Prospero et~al.(2002)Prospero, Ginoux, Torres, Nicholson, and Gill}]{prospero2002environmental}
Prospero, J.~M., Ginoux, P., Torres, O., Nicholson, S.~E., and Gill, T.~E.: Environmental characterization of global sources of atmospheric soil dust identified with the Nimbus 7 Total Ozone Mapping Spectrometer (TOMS) absorbing aerosol product, Reviews of geophysics, 40, 2--1, 2002.

\bibitem[{Rashki et~al.(2013)Rashki, Kaskaoutis, Goudie, and Kahn}]{rashki2013dryness}
Rashki, A., Kaskaoutis, D., Goudie, A.~S., and Kahn, R.: Dryness of ephemeral lakes and consequences for dust activity: the case of the Hamoun drainage basin, southeastern Iran, Science of the total environment, 463, 552--564, 2013.

\bibitem[{Rezazadeh et~al.(2013)Rezazadeh, Irannejad, and Shao}]{rezazadeh2013climatology}
Rezazadeh, M., Irannejad, P., and Shao, Y.: Climatology of the Middle East dust events, Aeolian Research, 10, 103--109, 2013.

\bibitem[{Shao and Dong(2006)}]{shao2006review}
Shao, Y. and Dong, C.: A review on East Asian dust storm climate, modelling and monitoring, Global and Planetary Change, 52, 1--22, 2006.

\bibitem[{Stein et~al.(2015)Stein, Draxler, Rolph, Stunder, Cohen, and Ngan}]{stein2015noaa}
Stein, A.~F., Draxler, R.~R., Rolph, G.~D., Stunder, B.~J., Cohen, M.~D., and Ngan, F.: NOAA’s HYSPLIT atmospheric transport and dispersion modeling system, Bulletin of the American Meteorological Society, 96, 2059--2077, 2015.

\bibitem[{Stocker(2014)}]{stocker2014climate}
Stocker, T.: Climate change 2013: the physical science basis: Working Group I contribution to the Fifth assessment report of the Intergovernmental Panel on Climate Change, Cambridge university press, 2014.

\bibitem[{Wang et~al.(2009)Wang, Jeong, and Mahowald}]{wang2009particulate}
Wang, C., Jeong, G.-R., and Mahowald, N.: Particulate absorption of solar radiation: anthropogenic aerosols vs. dust, Atmospheric Chemistry and Physics, 9, 3935--3945, 2009.

\end{thebibliography}

\end{document}